\RequirePackage{mathtools}
\RequirePackage{commath, amsmath, amssymb, amsfonts}
\documentclass[10pt, a4paper]{iopart}
\usepackage{xcolor}
\usepackage{physics}
\usepackage{graphicx}
\usepackage{booktabs}
\usepackage[normalem]{ulem}
\newcommand{\q}{\mathbf{q}}
\graphicspath{{fig/}}

\newcommand{\add}[1]{\textcolor{black}{#1}}
\begin{document}
\title{Exploring and machine learning structural instabilities in 2D materials}
\author{Simone Manti$^{\xi,*}$, Mark Kamper Svendsen$^{\xi}$, Nikolaj R. Knøsgaard$^{\xi}$, Peder M. Lyngby$^{\xi}$, and Kristian S. Thygesen$^{\xi,\dagger}$}
\address{$^\xi$CAMD, Computational Atomic-Scale Materials Design, Department of Physics, Technical University of Denmark, 2800 Kgs. Lyngby Denmark}
\address{$^{\dagger}$Center for Nanostructured Graphene (CNG),
Department of Physics, Technical University of Denmark, DK - 2800 Kongens Lyngby, Denmark}
\address{$^*$Corresponding author: simone.manti@lnf.infn.it}
\vspace{10pt}
\begin{abstract}
We address the problem of predicting the zero-temperature
dynamical stability (DS) of a periodic crystal without computing its full phonon band structure. Here we report the evidence that DS can be inferred with good reliability from the phonon frequencies at the center and boundary of the Brillouin zone (BZ). This analysis represents a validation of the DS test employed by the Computational 2D Materials Database (C2DB). For 137 dynamically unstable 2D crystals, we displace the atoms along an unstable mode and relax the structure. This procedure yields a dynamically stable crystal in 49 cases. The elementary properties of these new structures are characterised using the C2DB workflow, and it is found that their properties can differ significantly from those of the original unstable crystals, e.g. band gaps are opened by 0.3 eV on average. All the crystal structures and properties are available in the C2DB. Finally, we train a classification model on the DS data for 3295 2D materials in the C2DB using a representation encoding the electronic structure of the crystal. We obtain an excellent receiver operating characteristic (ROC) curve with an area under the curve (AUC) of 0.90, showing that the classification model can drastically reduce computational efforts in high-throughput studies.
\end{abstract}
\ioptwocol
\section{Introduction}
Computational materials discovery aims at identifying materials for specific applications, often employing first principles methods such as density functional theory (DFT) \cite{kohmsham}. The potential of a given material for the targeted application is usually evaluated based on elementary properties of the crystal, such as the electronic band gap, the optical absorption spectrum, or the magnetic order. Such properties can be highly sensitive to even small distortions of the lattice that reduce the symmetry of the crystal, and it is therefore important to develop efficient methods for identifying and accounting for such distortions.\par
Lattice distortions can be classified according to their periodicity relative to the primitive cell of the crystal. Local instabilities conserve the periodicity of the crystal, i.e. they do not enlarge the number of atoms in the primitive cell. Other distortions, \add{accompanied by a modulation of the electronic density} known as charge density wave (CDW) \cite{GrunerCDW}, lead to an enlargement of the period of the crystal, which can be either commensurate or incommensurate with the high-symmetry phase. \add{A classical example is the Peierls instability in one-dimensional  \cite{peierls1996quantum} and two-dimensional (2D) \cite{yoshiyuki@2000} systems, where a gap is opened in the CDW state.} A universal microscopic theory of the CDW phase is still missing due to the many possible and intertwined driving mechanisms, e.g. electron-phonon interaction \cite{elph@TiSe2}, Fermi surface nesting, or phonon-phonon interactions \cite{Anharmonic@Mauri}, which makes a precise clear-cut definition of the CDW phase difficult. In addition, the CDW state is sensitive to external effects such as temperature and doping \cite{AnharmonicityDoping@Mauri}. As a testimony to the complexity of the problem, different models and concepts are used to describe the CDW phase depending on the dimensionality of the material  \cite{dimensCDW,dimensCDW2,dimensCDW3,natureCDW,johannes2008fermi}.\par
The last few years have witnessed an increased interest in CDW states of \add{2D} materials. For example, CDW physics is believed to govern the transition from the trigonal prismatic T-phase to the lower symmetry T’-phase in monolayer MoS$_2$ \cite{TTpMoS2} as well as the plethora of temperature dependent phases in monolayers of NbSe$_2$ \cite{NbSe2CDW,NbSe2CDW2}, TaS$_2$  \cite{TaS2CDW,TaS2CDW2}, TaSe$_2$ \cite{TaSe2CDW,TaSe2CDW2}, and TiSe$_2$  \cite{TiSe2CDW,TiSe2CDW2}. In addition, a number of recent studies have investigated the possibility to control CDW phase transitions. For instance, the T-phase of monolayer MoS$_2$ can be stabilized by argon bombardment \cite{ArTMoS2}, exposure to electron beams \cite{TTpMoS2}, or Li-ion intercalation \cite{Li-inter-TMoS2}. Similar results have been reported for MoTe$_2$ \cite{MoTe2transition}.\par
Regardless of the fundamental origin of possible lattice distortions, it remains of great practical importance to devise efficient schemes that makes it possible to verify whether or not a given structure is dynamically stable (DS), i.e. whether it represents a local minimum of the potential energy surface. Structures that are not DS are frequently generated in computational studies, e.g. when a structure is relaxed under symmetry constraints or the chosen unit cell is too small to accommodate the stable phase. Tests for DS are rarely performed in large-scale discovery studies, because there is no established way of doing it apart from calculating the full phonon band structure \cite{Mounetphdb}, which is a time-consuming task. At the same time, the importance of incorporating such tests is in fact unclear; that is, it is not known how much symmetry-breaking distortions generally influence the properties of a materials.\par
A straightforward strategy to generate potentially stable structures from dynamically unstable ones, is to displace the atoms along an unstable phonon mode using a supercell that can accommodate the distortion. This approach has previously been adopted to explore structural distortions \add{in metallic system \cite{togo@2013}} in bulk perovskites  \cite{tempPatrick,walsh1} and one-dimensional organometallic chains  \cite{1dcdwdb}. However, systematic studies of structural instabilities in 2D materials, have so far been lacking.\par
In this work, we perform a systematic study of structural distortions across a broad class of 2D crystals, and explore a machine learning-based approach to DS classification. Throughout, we focus on the most common case of small-period, commensurate distortions that can be accommodated in a $2\times 2$ repetition of the primitive cell of the high-symmetry phase. We shall refer to the test for the occurrence of such distortions as the Center and Boundary Phonon (CBP) protocol. The motivation behind the present work is fourfold: (i) To assess the reliability of the CBP protocol (which is currently used for DS classification in the Computational 2D Materials Database (C2DB)\cite{Haastrup_2018,gjerding2021recent}). (ii) To elucidate the effect of symmetry-breaking distortions on the basic electronic properties of crystals. (iii) To obtain the DS phases of a set of dynamically unstable 2D materials that were originally generated by combinatorial lattice decoration, and make them available to the community via the C2DB. (iv) To explore the viability of a machine learning based classification scheme for predicting DS using input from a DFT calculation of the undistorted high-symmetry phase. 
\par
The  paper  is  structured as follows. In Section \ref{sec:methodology} we describe the CBP protocol. We first benchmark the CBP protocol against full phonon band structure calculations and evaluate its statistical success rate. For 137 dynamically unstable 2D materials, we further analyse how the small-period distortions that stabilise the materials influence their electronic properties. Section 3 concludes the paper. 
\section{\add{Results and discussion}}\label{sec:methodology}
\begin{figure*}[t]
    \centering
    \includegraphics[width=1.\linewidth]{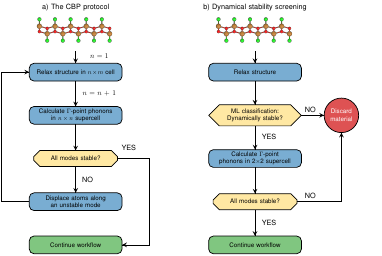}
    \caption{\textbf{Workflow diagram.} Workflow to test for dynamical stability of (two-dimensional) crystals in high-throughput computational studies. In a first step, a machine learning classification algorithm is used to filter out unstable crystals at minimal computational cost. The stability of the remaining crystals are assessed using the CBP protocol. Only materials predicted to be stable by the CBP protocol continue to the characterisation workflow.}
    \label{fig:cbp}
\end{figure*}

\add{This section presents and discusses the results of the CBP protocol and the connection with the machine learning predictions. Together they will increase the success rate of finding dynamically stable material within the C2DB workflow (see Figure (\ref{fig:cbp})).}
\begin{figure*}[h!]
    \centering
    \includegraphics[width=1.\textwidth]{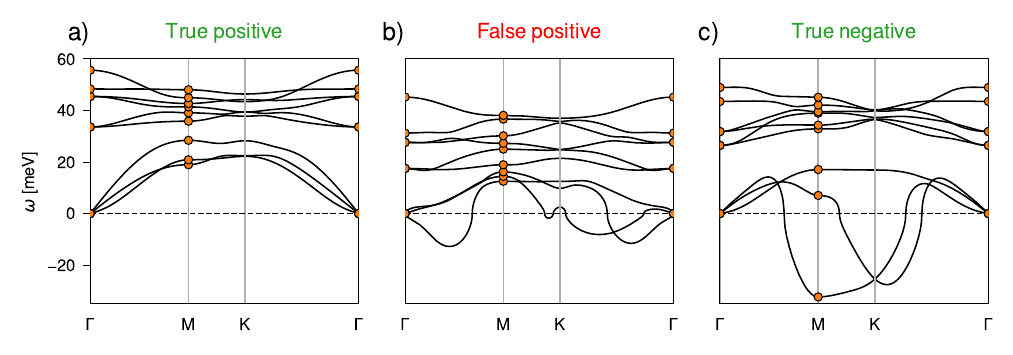}
    \caption{\textbf{Possible outcomes of the CBP protocol.} Phonon band structure for monolayer MoS$_2$ in the H-phase (a), NbSSe (b), and MoS$_2$ in the T-phase (c). Note that imaginary phonon frequencies are represented by negative values. The CBP protocol (orange dots) is sufficient to conclude that a material is dynamically stable (unstable) in the situations depicted in the (a) and (c) panels. In contrast, when the relevant distortion requires a supercell larger than a $2 \times 2$, and the phonon frequencies are real at the center and boundary of the BZ, the CBP protocol will result in a false positive result.}
    \label{fig:outprotocol}
\end{figure*}
\subsection{The CBP protocol: Stability test}
Given a material that has been relaxed in some unit cell (from hereon referred to as the primitive unit cell), the CBP protocol proceeds by evaluating the stiffness tensor of the material and the Hessian matrix of a supercell obtained by repeating the primitive cell $2\times 2$ times. In the current work, the stiffness tensor is calculated as a finite  difference of the stress under an applied strain, while the Hessian matrix is calculated as a finite difference of the forces on all the atoms of the $2\times 2$ supercell under displacement of the atoms in one primitive unit cell (this is equivalent to calculating the phonons at the center and specific high symmetry points at boundary of the BZ of the primitive cell, see Fig. (\ref{fig:hist}). Next, the stiffness tensor and the Hessian matrix are diagonalised, and the eigenvalues are used to infer a structural stability. A negative eigenvalue of the stiffness tensor indicates an instability of the lattice (the shape of the unit cell) while a negative eigenvalue of the $2\times 2$ Hessian signals an instability of the atomic structure. The obvious question here, is whether it suffices to consider the Hessian of the $2\times 2$ supercell, or equivalently consider the phonons at the BZ center and boundaries. \par
We can distinguish three possible outcomes when comparing the CBP protocol against full phonon calculations (see Figure (\ref{fig:outprotocol})), namely a true positive result, a true negative result, and a false positive result. We note that the case of a false negative is not possible, because a material that is unstable in a $2\times 2$ cell is de facto unstable.  The false positive case occurs when a material is stable in a $2\times 2$ supercell, but unstable if allowed to distort in a larger cell. Our results show that such large-period distortions that do not show as distortions in a $2\times 2$ cell, are relatively rare (see Section \ref{sec:benchmark}).

\subsection{The CBP protocol: Structure generation}\label{sec:generation}
\begin{figure*}[h!]
    \centering
    \includegraphics[width=1.\textwidth]{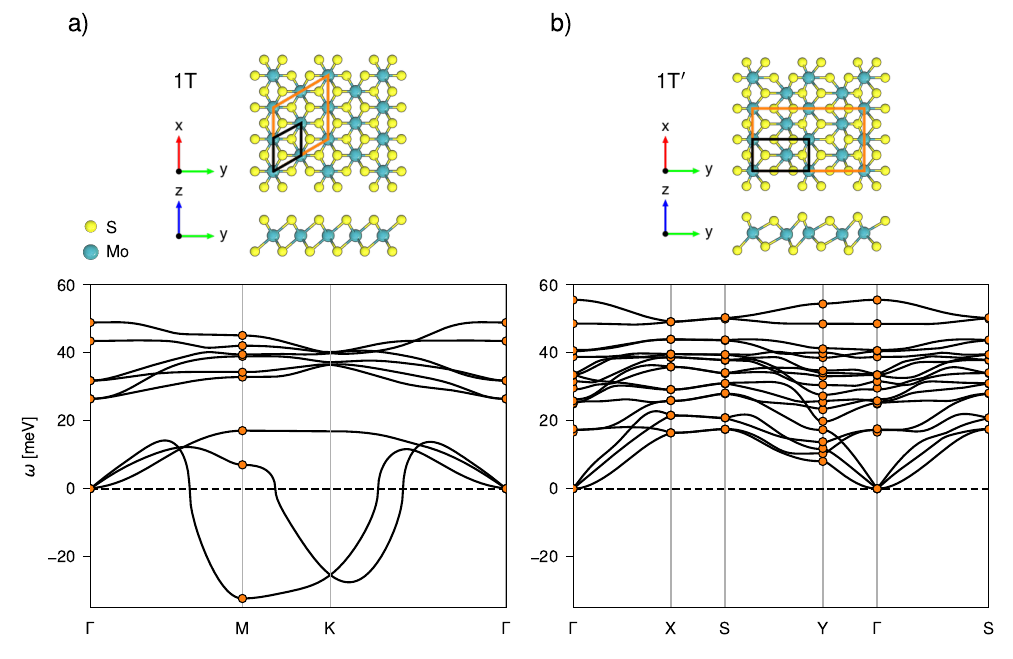}
    \caption{\textbf{Phonons frequencies of the T and T' phase of MoS$_2$.} The CBP protocol captures the instability of the T-phase (a) of MoS$_2$. Both the primitive unit cell (black) and the $2\times 2$ supercell (orange) are shown. Displacing the atoms along the unstable TA mode at the M-point ($\mathbf{q} = (\frac{1}{2},0)$), which can be accommodated in the $2\times 1$ supercell, and subsequently relaxing the structure results in the dynamically stable T'-phase (b). }
    \label{fig:mos2bands}
\end{figure*}

\begin{figure*}[h!]
    \centering
    \includegraphics[width=1.\textwidth]{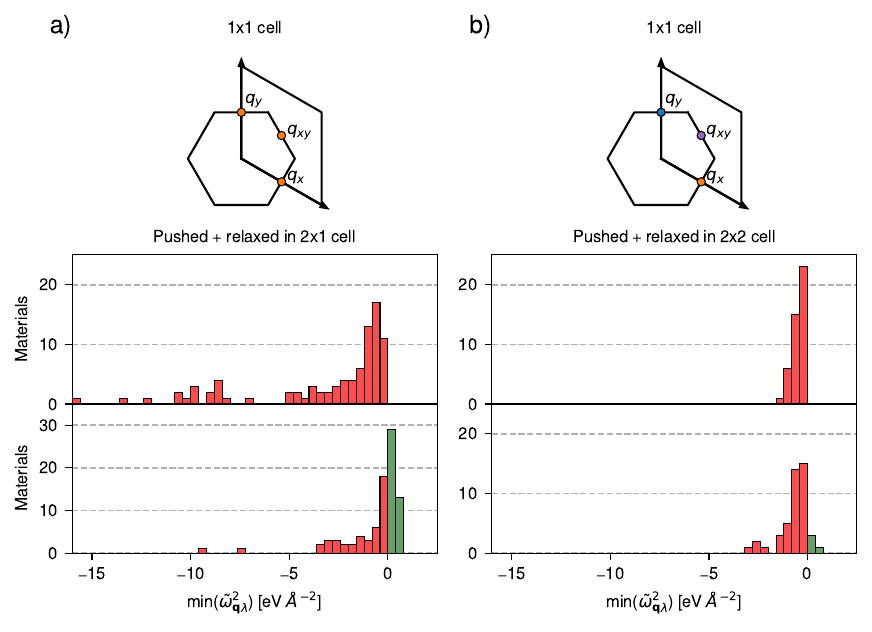}
    \caption{\textbf{Histograms of minimum eigenvalues of the Hessian matrix before and after displacing along the unstable mode.} The 137 dynamically unstable 2D materials studied in this work can be divided into two groups depending on whether the minimum eigenvalues of the Hessian matrix \add{$\tilde{\omega}_{\q\lambda}$} at $q=\{(\frac{1}{2},0),(0,\frac{1}{2}),(\frac{1}{2},\frac{1}{2})\}$ are equal (panel (a) orange) or different (panel (b) different colors). For the first group of materials, the atoms are displaced along the mode at $q=(\frac{1}{2},0)$ and relaxed in a $2\times 1$ supercell. \add{Starting from 91 unstable materials (red), 43 (green) become stable}. For the second group, displacing along $q=(\frac{1}{2},\frac{1}{2})$ and relaxing in a $2\times 2$ supercell, yields a dynamically stable structure in 6/46 cases.}
    \label{fig:hist}
\end{figure*}

Here we outline a simple procedure to generate distorted and potentially stable structures from an initial dynamically unstable structure. The basic idea is to displace the atoms along an unstable phonon mode followed by a relaxation. In practice, the unstable mode is obtained as the eigen function corresponding to a negative eigenvalue of the Hessian matrix of the $2\times 2$ supercell. The procedure is illustrated in Figure (\ref{fig:mos2bands}) for the well known T-T' phase transition of MoS$_2$ \cite{TTpMoS2}. The left panel shows the atomic structure and phonon band structure of monolayer MoS$_2$ in the T-phase. Both the primitive unit cell (black) and the $2\times 2$ supercell (orange) are indicated. The CBP method identifies an unstable mode at the BZ boundary (M point). After displacing the atoms along the unstable mode, a distorted structure is obtained, which after relaxation leads to the dynamically stable T'-phase of MoS$_2$ shown in the right panel.   
In this work, we have applied the CPB protocol systematically to 137 dynamically unstable 2D materials. The 137 monolayers were selected from the C2DB according to the following two criteria: First, to ensure that all materials are chemically "reasonable", only materials with a low formation energy were selected. Specifically, we require that  $\Delta H_\text{hull} < 0.2$ eV atom\textsuperscript{-1}, where $\Delta H_\text{hull}$ is the energy above the convex hull defined by the most stable (possibly mixed) bulk phases of the relevant composition \cite{C2DB,gjerding2021recent}. Secondly, we consider only materials with exactly one unstable mode, i.e. one negative eigenvalue of the Hessian matrix at a given $q$-point. \add{We stress that the latter condition is not strictly necessary but was adopted here to limit the number of materials. When two or more unstable modes exist there is not a unique way to distort the structure. One possibility is to push along the linear combination of modes yielding the distorted structure with the highest symmetry \cite{togo@2013}. However, it is not clear that imposing high symmetry is the best strategy for finding a DS structure. For the few cases with multiple unstable modes that we have analysed, we have found that pushing along the most unstable eigenmode (the one with the most negative eigenvalue) often yields a DS structure, like in the case of T-MoS\textsubscript{2}.}

For the 137 dynamically unstable materials we displaced the atoms along the (unique) unstable mode. The size of the displacement was chosen such that the maximum atomic displacement was exactly 0.1 \AA. This displacement size was chosen based on the MoS$_2$ example discussed above, where it results in a minimal number of subsequent relaxation steps. A smaller value does not guarantee that the system leaves the saddle point, while a larger value creates a too large distortion resulting in additional relaxation steps. During relaxation the unit cell was allowed to change with no symmetry constraints and the relaxation was stopped when the forces on all atoms were below 0.01 eV \AA\textsuperscript{-1}.
\subsection{Assessment of the CBP protocol}\label{sec:benchmark}
To test the validity the CBP protocol, we have performed full phonon calculations for a set of 20 monolayers predicted as dynamically stable by the CBP protocol. The 20 materials were randomly selected from the C2DB and cover 7 different crystal structures. Out of the 20 materials 10 are metals and 10 are insulators/semiconductors. The calculated phonon band structures are reported in the \add{Supplementary information} (SI). For all materials, the phonon frequencies obtained with the CBP protocol equal the frequencies of the full phonon band structure at the $q$-points $\mathbf{q} \in \{(0,0),(\frac{1}{2},0), (0,\frac{1}{2}), (\frac{1}{2},\frac{1}{2})\}$. This is expected as the phonons at these $q$-points can be accommodated by the $2\times 2$ supercell. 

Within the set of 20 materials, we find three False-positive cases, namely CoTe$_2$, NbSSe, and TaTe$_2$. These materials exhibit unstable modes (imaginary frequencies or equivalently negative force constant eigenvalues) in the interior of the BZ (NbSSe and TaTe$_2$) or at the K-point (CoTe$_2$), while all phonon frequencies at the $q$-points covered by the CBP protocol, are real. \add{A simple interpolation of the frequencies is not enough to catch an instability at internal points of BZ (see  Supplementary Figure 1) because a supercell larger than a 2x2 is needed to catch these distortions, like in the case of 2H-CoTe2 (see  Supplementary Figure 2).} This relatively low percentage of False-positives in our representative samples is consistent with the work by Mounet et al. \cite{Mounetphdb} who computed the full phonon band structure of 258 monolayers predicted to be (easily) exfoliable from known bulk compounds. Applying the CBP protocol to their data yields 14 False-positive cases; half of these are transition metal dichalcogenides (TMDs) with Co, Nb or Ta. \par

We note that the small imaginary frequencies in the out of plane modes around the $\Gamma$-point seen in some of the phonon band structures are not distortions, but are rather due to the interpolation of the dynamical matrix. In particularly, these artifacts occur because of the broken crystal point-group symmetry in the force constant matrix and they will vanish if a larger supercell is used or the rotational sum rule is imposed \cite{hiphive,carrete@2016} \add{or higher-order multipolar interactions are included \cite{royo@2020}}.

\subsection{Stable distorted monolayers}
\begin{table*}[h!]
    \centering
    \footnotesize
    \begin{tabular}{l cc cc cc cc}
\toprule
Material & 
\multicolumn{2}{c}{Space group - Wyckoff} &
\multicolumn{2}{c}{$\Delta$H$_\text{hull}$ [eV atom\textsuperscript{-1}]} &
\multicolumn{2}{c}{\add{min($\tilde{\omega}_{q\lambda}^2$)} [eV \AA\textsuperscript{-2}]} &
\multicolumn{2}{c}{$\varepsilon_\textup{gap}^{\text{\tiny PBE}}$ [eV]} \\
\cmidrule(lr){2-3} \cmidrule(lr){4-5} \cmidrule(lr){6-7} \cmidrule(lr){8-9}
& before & after & before & after & before & after & before & after \\ 
\midrule
AgBr$_{2}$  & \add{164-bd}  & \add{14-ae}  &  0.05 & 0.00 & -1.01  & 0.00 & 0.00 & 0.00 \\
AgCl$_{2}$  & 164-bd  & \add{14-ae}  &  0.05 & 0.00 & -2.60  & 0.06 & 0.00 & 0.00 \\
AsClTe      & 156-ac  & \add{156-abc}  &  0.19 & 0.02 & -0.51  & 0.25 & 1.29 & 1.48 \\
CdBr        & \add{164-ad}     & \add{13-d}  &  0.18 & 0.07 & -1.00  & 0.03 & 0.00 & 1.28 \\
CdCl        & \add{164-d}  & \add{164-cd}  &  0.17 & 0.04 & -0.74  & 0.17 & 0.00 & 1.67 \\
CoSe        & 164-bd  & \add{25-acgh}  &  0.05 & 0.03 & -4.82  & 0.09 & 0.00 & 0.00 \\
CrBrCl      & 156-abc & \add{7-a}  &  0.11 & 0.06 & -0.98  & 0.15 & 0.00 & 0.64 \\
CrBr$_{2}$  & 164-bd  & \add{14-de}  &  0.10 & 0.06 & -0.59  & 0.07 & 0.00 & 0.49 \\
CrCl$_{2}$  & 164-bd  & \add{14-be}  &  0.11 & 0.05 & -1.87  & 0.15 & 0.00 & 0.76 \\
CrSSe       & 156-abc & \add{6-ab}  &  0.15 & 0.09 & -9.75  & 0.71 & 0.00 & 0.00 \\
CrS$_{2}$   & \add{164-bd} & \add{11-e}  &  0.18 & 0.05 & -15.62 & 0.74 & 0.00 & 0.00 \\
CrSe$_{2}$  & \add{164-bd} & \add{11-e}  &  0.14 & 0.05 & -12.17 & 0.77 & 0.00 & 0.00 \\
CrTe$_{2}$  & \add{164-bd} & \add{11-e}  &  0.02 & 0.01 & -1.71 & 0.08 & 0.00 & 0.00 \\
CrPS$_{3}$  & \add{162-dek}   & \add{11-ef}  &  0.09 & 0.03 & -3.13 & 0.00 & 0.00 & 0.34 \\
FePSe$_{3}$ & \add{162-dek}  & \add{11-fe}  &  0.13 & 0.12 & -0.42 & 0.04 & 0.13 & 0.13 \\
FeSe$_{2}$  & \add{187-bi}  & \add{187-eg}  &  0.15 & 0.00 & -2.04 & 0.00 & 0.00 & 0.00 \\
HfBrCl      & 156-abc & \add{6-ab}  &  0.14 & 0.03 & -9.61 & 0.40 & 0.00 & 0.82 \\
HfBrI       & 156-abc & \add{6-ab}  &  0.22 & 0.05 & -10.41 & 0.39 & 0.00 & 0.73 \\
HfBr$_{2}$  & 164-bd  & \add{11-e}  &  0.14 & 0.04 & -10.21 & 0.41 & 0.00 & 0.8 \\
HfCl$_{2}$  & 164-bd  & \add{11-e}  &  0.14 & 0.04 & -8.51 & 0.38 & 0.00 & 0.85 \\
HgSe        & \add{164-d}  & \add{10-mn}  &  0.11 & 0.04 & -0.83 & 0.04 & 0.08 & 0.37 \\
HgTe        & \add{164-d}  & \add{10-mn}  &  0.11 & 0.04 & -0.61 & 0.04 & 0.08 & 0.37 \\
InTe        & \add{164-d}  & \add{129-c}  &  0.18 & 0.10 & -0.41 & 0.23 & 0.00 & 0.00 \\
InBrSe      & 59-ab   & \add{59-ab}  &  0.03 & 0.02 & -0.50 & 0.04 & 1.23 & 1.23 \\
InSe        & 187-hi  & \add{187-hi} &  0.00 & 0.00 & -0.28 & 0.01 & 1.39 & 1.39 \\
MoSeTe      & 156-abc & \add{6-ab}  &  0.20 & 0.08 & -10.67 & 0.69 & 0.00 & 0.00 \\
MoTe$_{2}$  & 164-bd  & \add{11-e}  &  0.17 & 0.01 & -13.23 & 0.67 & 0.00 & 0.00 \\
NbS$_{2}$   & 187-bi  & \add{187-eg} &  0.00 & 0.00 & -1.08 & 0.06 & 0.00 & 0.00 \\
NbTe$_{2}$  & 187-bi  & \add{187-fg}  &  0.00 & 0.00 & -0.37 & 0.51 & 0.00 & 0.00 \\
PdI$_{2}$   & 164-bd  & \add{14-ae}  &  0.17 & 0.03 & -0.56 & 0.12 & 0.00 & 0.59 \\
RhI$_{2}$   & 164-bd  & \add{164-ad}  &  0.17 & 0.17 & -0.64 & 0.04 & 0.00 & 0.00 \\
RhO$_{2}$   & 164-bd  & \add{11-e}  &  0.16 & 0.15 & -2.19 & 0.65 & 0.00 & 0.00 \\
RhTe$_{2}$  & 164-bd  & \add{11-e}  &  0.11 & 0.07 & -1.67 & 0.36 & 0.00 & 0.13 \\
ScI$_{3}$   & 162-dk  & \add{14-e}  &  0.00 & 0.00 & -0.25 & 0.02 & 1.85 & 1.85 \\
TiBrCl      & \add{156-ac} & \add{6-ab}  &  0.05 & 0.00 & -9.15 & 0.52 & 0.00 & 0.29 \\
TiBr$_{2}$  & 164-bd  & \add{11-e}  &  0.06 & 0.04 & -0.54 & 0.30 & 0.00 & 0.12 \\
TiCl$_{2}$  & 164-bd  & \add{11-e}  &  0.11 & 0.00 & -9.99 & 0.53 & 0.00 & 0.32 \\
TiO$_{2}$   & 164-bd  & \add{11-e} &  0.14 & 0.12 & -1.96 & 0.47 & 2.70 & 2.85 \\
TiS$_{2}$   & 187-bi  & \add{31-a}  &  0.14 & 0.14 & -0.49 & 0.00 & 0.73 & 0.79 \\
TiPSe$_{3}$ & \add{162-dek} & \add{2-i}  &  0.16 & 0.00 & -1.24 & 0.00 & 0.00 & 0.00 \\
VTe$_{2}$   & 164-bd  & \add{11-e} &  0.02 & 0.00 & -1.05 & 0.38 & 0.00 & 0.00 \\
ZrBrCl      & 156-abc & \add{6-ab}  &  0.12 & 0.02 & -8.05 & 0.40 & 0.00 & 0.59 \\
ZrBrI       & 156-abc & \add{6-ab}  &  0.15 & 0.02 & -8.89 & 0.38 & 0.00 & 0.48 \\
ZrBr$_{2}$  & 164-bd  & \add{11-e}  &  0.11 & 0.00 & -8.65 & 0.47 & 0.00 & 0.59 \\
ZrClI       & 156-abc & \add{6-ab}  &  0.19 & 0.07 & -8.69 & 0.26 & 0.00 & 0.48 \\
ZrCl$_{2}$  & 164-bd  & \add{11-e}  &  0.11 & 0.03 & -7.13 & 0.46 & 0.00 & 0.60 \\
ZrI$_{2}$   & 164-bd  & \add{11-e}  &  0.14 & 0.00 & -8.59 & 0.48 & 0.00 & 0.43 \\
ZrS$_{2}$   & 187-bi  & \add{31-a}  &  0.19 & 0.18 & -0.80 & 0.15 & 0.96 & 1.13 \\
\bottomrule
\end{tabular}
    \caption{\textbf{Properties of the stable materials.} Some of the calculated properties of the subset of the 137 materials that became dynamically stable after displacing the atoms along an unstable phonon mode. The properties are shown before and after the distortion, i.e. for the original dynamically unstable structures and the final dynamically stable structures, respectively.}
    \label{table:1set}
\end{table*}
The 137 dynamically unstable materials, which were selected from the C2DB according to the criteria described in Section \ref{sec:generation}, can be divided into two groups depending on whether the eigenvalues of the Hessian at the wave vectors $q_x=(\frac{1}{2},0)$, $q_y=(0,\frac{1}{2})$ and $q_{xy} = (\frac{1}{2},\frac{1}{2})$, are equal or not. Equality of the eigenvalues implies an isotropic Hessian. For such materials, we generate distorted structures by displacing the atoms along the unstable mode at $q_x=(\frac{1}{2},0)$, followed by relaxation in a $2\times 1$ supercell.
\add{In the case of an anisotropic Hessian, in general, it may exist a particular combination of q-vectors that stabilizes the system. Here, we were interested in finding a general method to generate DS structures in a high-throughput way and therefore, we decided to displace only at $q_{xy} = (\frac{1}{2},\frac{1}{2})$ in a $2\times 2$ supercell.}\par
After atomic displacement and subsequent relaxation, the CBP protocol was applied again to test for dynamical stability of the distorted structures. Histograms of the minimum eigenvalue, min($\tilde\omega_{\q\lambda}^2$), of the Hessian matrix \add{at $\q$ for the unstable mode $\lambda$}, are shown in Figure (\ref{fig:hist}) with the materials before and after atomic displacement shown in the upper and lower panels, respectively. Negative eigenvalues, corresponding to unstable materials, are shown in red while positive eigenvalues are shown in green. \add{We removed the three translational modes with eigenvalues close to zero before extracting min($\tilde\omega_{\q\lambda}^2$).}
Out of the 137 unstable materials, 49 become dynamically stable (according to the CBP protocol). By far the highest success rate for generating stable crystals was found for the isotropic materials (left panel), where 43 out of 91 materials became stable while only 6 out of the 43 anisotropic materials became stable. \add{In principle, the procedure can be applied many times to increase the number of the DS structures, while here, we applied the protocol only once for computational reasons. An example where applying the protocol two times is the case of the 1T-TiSe\textsubscript{2} monolayer in the SI. In that case, the material is first displaced at $q_{xy} = (\frac{1}{2},\frac{1}{2})$ in a $2\times 2$ supercell. Then the unstable material obtained is again displaced along one of the two degenerate modes at $\Gamma$ and the final structure with the experimentally known CDW state \cite{TiSe2CDW} is obtained.} 

A wide range of elementary properties of the 49 distorted, dynamically stable materials were computed using the C2DB workflow (see Table 1 in  \cite{gjerding2021recent} for a complete list of the properties). The atomic structures together with the calculated properties are available in the C2DB. Table \ref{table:1set} provides an overview of the symmetries, minimal Hessian eigenvalues, total energies, and electronic band gap of the 49 materials before and after the distortion. 
\par
\begin{figure}[t]
    \centering
    \includegraphics[width=1.\linewidth]{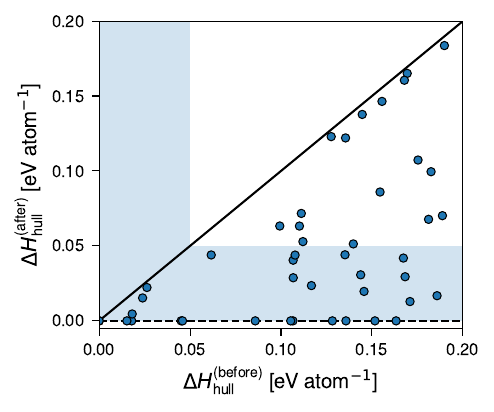}
    \caption{\textbf{Energy above the convex hull before and after the stabilization.}The energy above the convex hull for the 49 monolayers before and after distortion. Materials with a $\Delta H_\text{hull}^\text{(after)}$ close to zero are expected to be thermodynamically stable. The range up to 0.05 eV atom\textsuperscript{-1} above the convex hull has been indicated by a shaded blue region to visualise the importance of structural distortions for assessing the thermodynamic stability.}
    \label{fig:hull}
\end{figure}

Apart from the reduction in symmetry, the distortion also lowers the total energy of the materials. An important descriptor for the thermodynamic stability of a material is the energy above the convex hull, $\Delta H_{\mathrm{hull}}$. Figure \ref{fig:hull} shows a plot of $\Delta H_{\mathrm{hull}}$ before and after the distortion of the 49 materials. The reduction in energy upon distortion ranges from 0 to 0.2 eV atom\textsuperscript{-1}. In fact, several of the materials come very close to the convex hull and some even fall onto the hull, indicating their global thermodynamic stability (at $T=0$ K) with respect to the reference bulk phases. We note that all DFT energies, including the reference bulk phases, were calculated using the PBE xc-functional, which does not account for van der Waals interactions. Accounting for the vdW interactions will downshift the energies of layered bulk phases and thus increase $\Delta H_{\mathrm{hull}}$ for the monolayers slightly. This effect will, however, not influence the relative stability of the pristine and distorted monolayers, which is the main focus of the current work. 
\begin{figure}[h!]
    \centering
    \includegraphics[width=1.\linewidth]{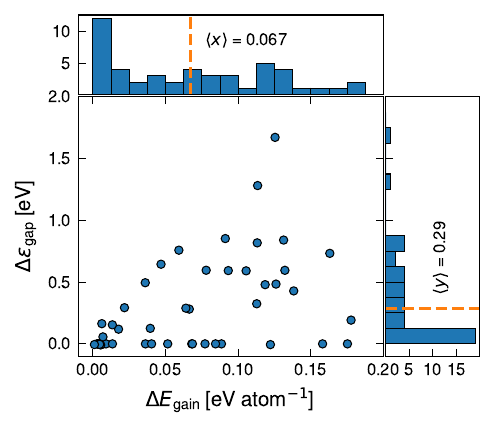}
    \caption{\textbf{The energy gain as a function the gap opening for the stable materials.} The average energy gain for the stable materials is \add{0.067} eV atom\textsuperscript{-1} and the average gap opening is 0.29 eV.}
    \label{fig:deltagap}
\end{figure}
Another characteristic trend observed is the opening/increase of the electronic band gap. The increase of the  single-particle band gap is expected to be related to the total energy gained by making the distortion. Figure \ref{fig:deltagap} shows the relation between the two quantities. Simplified models, for low dimensional systems and weak electron-phonon coupling, predict a proportionality between these two quantities \cite{Rossnagel_2011}. From our results it is clear that there is no universal relationship between the change in band gap and total energy. In particular, several of the metals show large gain in total energy while the gap remains zero.

\begin{figure*}[h!]
    \centering
    \includegraphics[width=1.\textwidth]{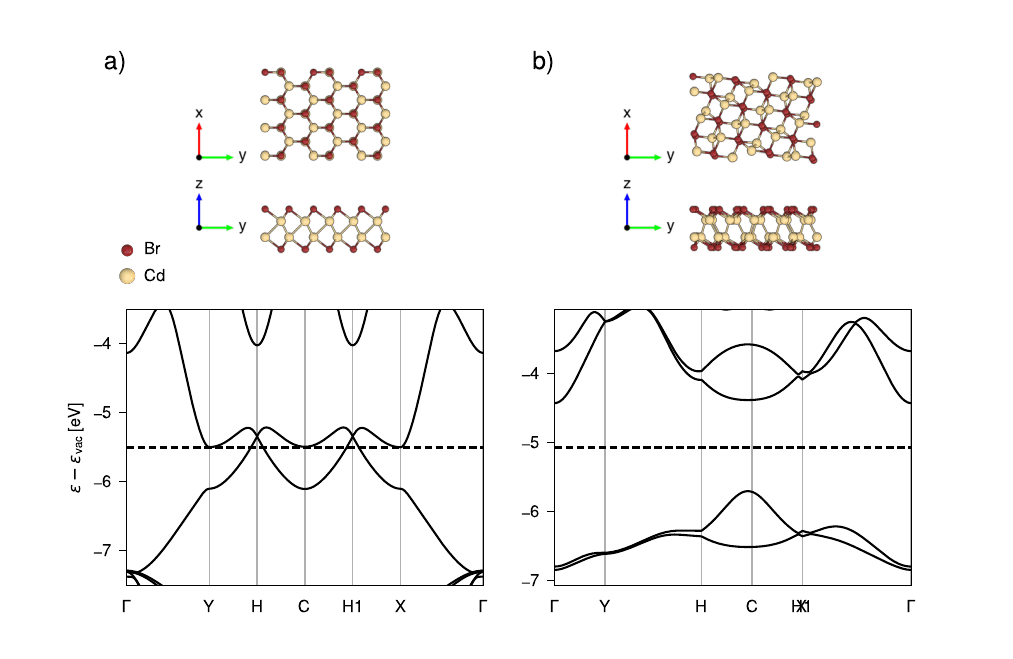}
    \caption{\textbf{Gap opening for CdBr.} Extreme case of gap opening for the stable material (CdBr) where the difference in the gap between the initial unstable metallic phase (a) and the final structure (b) is 1.28 eV.}
    \label{fig:matgap}
\end{figure*}

It is interesting that 21 of the distorted and dynamically stable materials exhibit direct band gaps when a tolerance of 0.1 eV is employed for the difference between the direct and indirect gap. Atomically thin direct band gap semiconductors are highly relevant as building blocks for opto-electronic devices, but only a hand full of such materials are known to date e.g. monolayers of the Mo- and W-based transition metal dichalcogenides \cite{mak2010atomically,manzeli20172d} and black phosphorous \cite{liu2014phosphorene}. As an example of a monolayer material that drastically changes from a metal to a direct band gap semiconductor upon distortion, we show the band structure of CdBr in Figure \ref{fig:matgap}. The initial unstable metallic phase of the material becomes dynamically stable upon distortion and opens a direct band gap of 1.28 eV at the C point.
\subsection{Machine learning dynamical stability}
\begin{figure*}[t]
    \centering
    \includegraphics[scale=1]{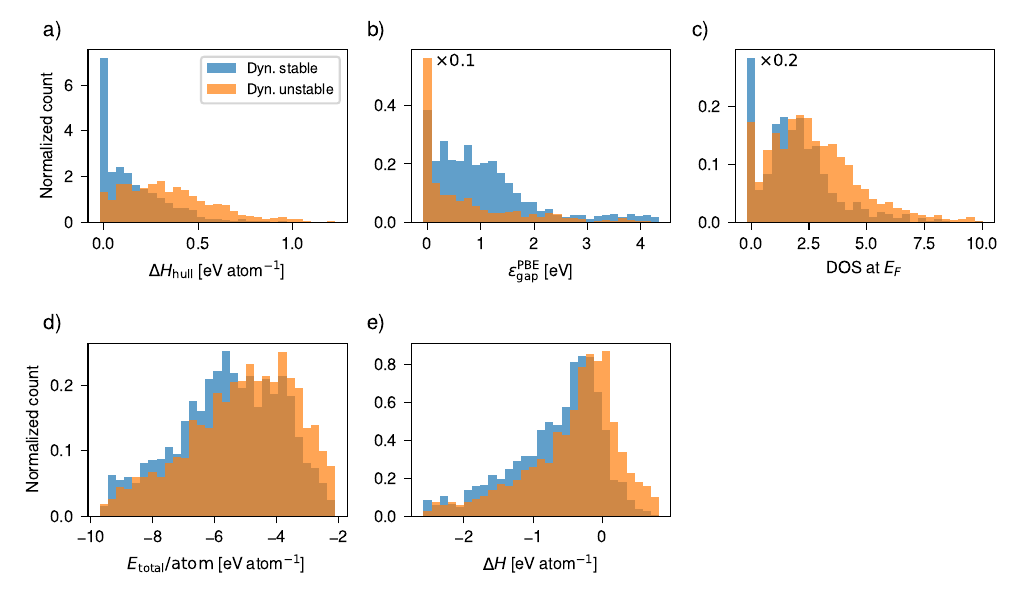}
    \caption{\textbf{Histograms of electronic features for stable and unstable materials.} Panel a) shows the distribution of the energy above the convex hull for high and low stability predicted materials. The stable materials tend to be closer to the convex hull. b) shows that materials predicted to be stable more frequently have a PBE band gap larger than zero. c) shows the DOS at the Fermi level distributions, which supports that stable materials have lower DOS at $E_F$. d) is for total energy per atom, which shows a slightly better separation between stable and unstable materials as stable materials tend to have a slightly lower total energy per atom. In e) the unstable materials tend to have a slightly larger heat of formation. Note that for b) and c) the peaks at $x=0$ for both stable and unstable materials have been scaled down by a factor of 10 and 5, respectively.}
    \label{fig:ml_hist}
\end{figure*}

\begin{figure*}[t]
    \centering
    \includegraphics[scale=1]{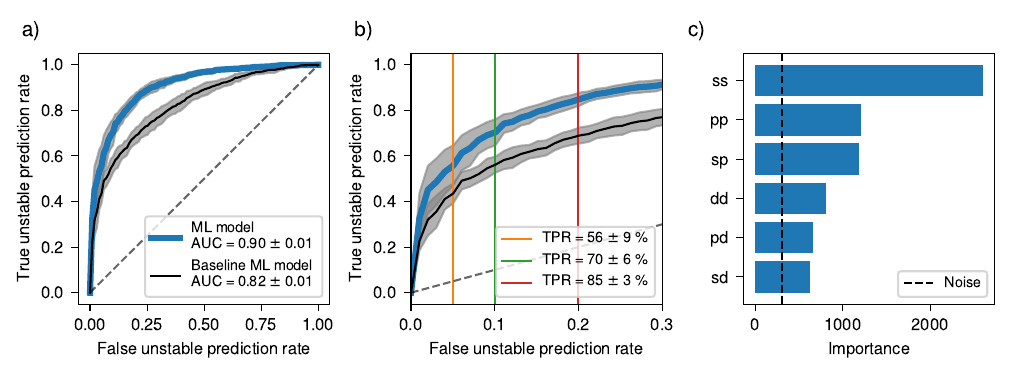}
    \caption{\textbf{Machine learning results.} Panel a) shows the ROC curves for a machine learning model trained on RAD-PDOS fingerprints and a baseline model trained on electronic features from the C2DB. The AUC scores are 0.91 and 0.82, respectively. b) shows the ROC curves zoomed on low false prediction rates with different classification thresholds highlighted with vertical lines. By accepting $5\%$ of the stable materials being falsely characterised as unstable, we can correctly label $56 \pm 9\%$ of the unstable materials. For $10\%$ the true prediction rate is $70 \pm 6\%$ and for $20\%$ it is $85 \pm 3\%$. c) shows feature importances of the different RAD-PDOS fingerprints measured as how many times the fingerprints are used to perform a split in the ML model. The RAD-PDOS ss fingerprint is found to be the most important fingerprint for the ML model.}
    \label{fig:ml_results}
\end{figure*}



We next attempt to accelerate the dynamic stability prediction using the machine learning model outlined in Section \ref{sec:ML}. 

As an introductory exercise we consider the correlation between dynamical stability and \add{five} elementary materials properties, namely the energy above the convex hull, the PBE band gap, the DOS at the Fermi level, the total energy per atom, and heat of formation. Figure \ref{fig:ml_hist} shows the distribution of these properties over the 3295 2D materials. The materials have been split into dynamically stable (blue) and dynamically unstable materials (orange), respectively. There is a clear correlation between dynamical stability and the first three materials properties shown in panels a-c. In particular, dynamically stable materials are closer to the convex hull, have larger band gap, and lower DOS at $E_F$ as compared to dynamically unstable materials. The observed correlation with $\Delta H_{\mathrm{hull}}$ is consistent with previous findings based on phonon calculations \cite{gjerding2021recent}. In contrast, no or only weak correlation is found for the last \add{two} quantities in panels d-e. These  \add{five} properties were used as a low-dimensional feature vector  for training an XGBoost machine learning model that will serve as a baseline for a model trained on the higher dimensional RAD-PDOS representation described in Sec. \ref{sec:ML}.

To evaluate the performance of our model we employ the receiver operating characteristic (ROC) curve. The ROC curve maps out the number of materials correctly predicted as unstable as a function of the number of materials incorrectly labelled as unstable, and it is calculated by varying the classification tolerance of the model. The area under the curve(AUC) is a measure of the performance of the classifier. Random guessing would amount to a linear ROC curve with unit slope, shown in Fig. \ref{fig:ml_results} by the dashed grey line, and correspond to an AUC of 0.5 whereas a perfect classification model would have an AUC of 1. When calculating the ROC curve of our dynamic stability classifier we employ ten fold cross-validation (CV). This allows us to obtain a mean ROC curve and its standard deviation, which we then use to evaluate the performance of our model. 

The results from the machine learning model is shown in Fig. \ref{fig:ml_results}. The mean ROC curve is shown in blue in Fig.\ref{fig:ml_results}a; it achieves an excellent 10-fold CV AUC of $0.90\pm0.01$. This suggests that the XGBoost model is able to efficiently detect the dynamically unstable materials in the C2DB. We quantify the effect of the RAD-PDOS fingerprints by comparing the performance of the full model with a model trained on \add{the low-dimensional fingerprint}. We observe that the effect on including the RAD-PDOS in the fingerprint is statistically significant raising the AUC from $0.82\pm 0.01$ to $0.90\pm 0.01$. The relative impact of the RAD-PDOS fingerprints is shown in the feature importance evaluation in Fig. \ref{fig:ml_results}c. Here feature importance refers to how many times a feature is used to perform a split in the decision trees, and the feature importance have been summed for the six different components of the RAD-PDOS fingerprints, i.e. summing the radial distance and energy axes of the fingerprint. The vertical dashed black line shows the feature importance of random noise for reference. We observe that especially the RAD-PDOS $ss$ fingerprint leads to many splits in the gradient boosted trees. \add{Part of the importance is explained by the number of non-zero features in the fingerprints. For the materials without $d$-orbital electrons, the $sd$, $pd$ and $dd$ fingerprints will be all zero, while the $ss$, $sp$ and $pp$ will have fewer zero-valued features and thus more features to use for splits in the trees.}

Because of the strong performance of the model, we envision that it can be deployed directly after the initial relaxation step of a high-throughput workflow to reduce the number of phonon calculations needed to remove the dynamically unstable materials \add{as depicted in Figure \ref{fig:cbp}. The ML model does not replace the phonon calculation but is merely used to avoid performing expensive phonon calculations for materials that can be labelled unstable by the ML model}. Depending on the number of stable candidates that one is willing to falsely label as unstable, it is possible to save a significant amount of phonon-calculations by pre-screening with the ML model. The willingness to sacrifice materials is controlled by the classification tolerance. The trade-off between the number of unstable materials removed and the number of stable materials lost is directly mapped out by the ROC curve. In Fig. \ref{fig:ml_results}b we have indicated the classification thresholds where we \add{discard} 5\%, 10\% and 20 \% of the stable materials, and we observe that we can save $56\pm 9\%$, $70\pm 6\%$ and $85\pm 3\%$ of the computations for the three thresholds, respectively. \\

As an additional test of the machine learning model we apply it to the set of the 137 dynamically unstable materials that were investigated using the CBP protocol in the first part of the paper. The dynamical stability of the materials is evaluated by the ML model both before and after being pushed along an unstable mode (recall that before the push all the 137 materials are unstable; after the push the subset of 49 materials listed in Table 1 become stable while the other materials remain unstable). It is found that before the push $56\%$ of the unstable materials are labeled correctly. After the push, only $29\%$ of the unstable materials are labelled as unstable while the precision of the stable materials are $72\%$. Overall, the ML model performs worse on this test set than on a randomly selected test set from the original data set. An obvious explanation is that the 137 materials were selected according to (i) low energy above the convex hull ($\Delta H_{\mathrm{hull}}<0.2\, \mathrm{eV}/\mathrm{atom}$) and (ii) dynamically unstable. As seen from Fig. \ref{fig:ml_hist}a such materials are highly unusual and not well represented by the set of materials used to train the model.

%
In conclusion, we have performed a systematic study of structural instabilities in 2D materials. We have validated a simple protocol (here referred to as the CBP protocol) for identifying dynamical instabilities based on the frequency of phonons at the center and boundary of the BZ. The CBP protocol correctly classifies 2D materials as dynamically stable/unstable in 236 out of 250 cases \cite{Mounetphdb} and is ideally suited for high-throughput studies where the computational cost of evaluating the full phonon band structure becomes prohibitive. 

For 137 dynamically unstable monolayers with low formation energies, we displaced the atoms along an unstable phonon mode and relaxed the structure in a $2\times 1$ or $2\times 2$ supercell. This resulted in 49 distorted, dynamically stable monolayers. The success rate of obtaining a dynamically stable structure from this protocol was found to be significantly higher for materials with only one unstable phonon mode as compared to cases with several modes. In the latter case, the displacement vector is not unique and different choices generally lead to different, (dynamically unstable) structures. The 49 stable structures were fully characterised by an extensive computational property workflow and the results are available via the C2DB database. The properties of the distorted structures can deviate significantly from the original high symmetry structures, and we found only a weak, qualitative relation between the gain in total energy and band gap opening upon distortion.  
Finally, we trained a machine learning classification model to predict the dynamical stability using a radially decomposed projected density of states (RAD-PDOS) representation as input and a gradient boosting decision tree ensemble method (XGBoost) as learning algorithm. The model achieves an excellent ROC-AUC score of $0.90$ and lends itself to high-throughput assessment of dynamical stability. 
\section{\add{Methods}}
\subsection{\add{Density functional theory calculations}}
All phonon calculations were performed using the \texttt{asr.phonopy} recipe of the Atomic Simulation Recipes (ASR) \cite{gjerding2021asr}, which makes use of the Atomic Simulation Environment (ASE) \cite{larsen2017atomic} and PHONOPY \cite{TOGO20151}. The DFT calculations were performed with the GPAW \cite{Enkovaara_2010} code and the Perdew-Burke-Ernzerhof (PBE) exchange-correlation functional  \cite{PhysRevLett.77.3865}. The BZ was sampled on a uniform $k$-point mesh of density of 6.0 \AA$^2$ and the plane wave cutoff was set to 800 eV. To evaluate the Hessian matrix, the small displacement method was used  with a displacement size of 0.01 \AA$\,$ and forces were converged up to 10$^{\textrm{-4}}$ eV \AA\textsuperscript{-1}. \add{The non-analytical force constants were not included because we saw that they do not have any effect on the minimum eigenvalues of the hessian.} To benchmark the CBP protocol, we compare to full phonon band structures. In these calculations, the size of the supercell is chosen such that the Hessian matrix includes interactions between pairs of atoms within a radius of at least 12 \AA. (This implies that the supercell must contain a sphere of radius 12 \AA). \add{The spacegroup of the materials before and after applying the protocol, were calculated with spglib \cite{spglib2009togo}, with a symmetry threshold of 0.1 \AA, a value also used in a similar study \cite{togo@2013}.}
\subsection{\add{Machine learning method}}\label{sec:ML}
Below we describe the machine learning algorithm that we have developed and assessed in an attempt to accelerate the prediction of dynamic instabilities. Our choice of machine learning algorithm is the library, XGBoost \cite{chen_xgboost}, due to its robustness and flexibility, while being a simpler model compared to neural network methods. XGBoost is a regularized high-performance implementation of gradient tree boosting, which makes predictions based on an ensemble of gradient boosted decision trees. The decision trees of the ensemble are grown sequentially while learning from the mistakes of the previous trees by minimizing the loss function through gradient descent. This loss function is regularized to reduce the complexity of the individual decision trees which reduces the risk of overfitting. In contrast, in the widely used decision tree ensemble model Random Forest, the decision trees are grown independently and without any regularization. \\
The dataset used is a subset of C2DB and consists of 3212 materials (1536 stable and 1676 unstable materials), which does not include the 137 distorted materials identified in the first part of the paper (as these will be used as a particularly challenging test case for the model performance). As input for the model we use the radially projected density of states (RAD-PDOS) material fingerprints \cite{Knosgaard2022}.
The RAD-PDOS starts from the wave functions projected onto the atomic orbitals ($\nu$) of all the atoms ($a$) of the crystal, $\rho_{nk}^{a\nu} = |\langle \psi_{nk}  | a\nu\rangle|^2$. For each state, these projections are then combined into a radially distributed orbital pair correlation function, 
\begin{align}
    \rho_{nk}^{\nu \nu'}(R) &= \sum_{aa'} \rho_{nk}^{a\nu} \rho_{nk}^{a'\nu'} G\left(R - |R_{a}-R_{a'}|;\delta_R\right) \nonumber\\ &\times \exp \left(-\alpha_R R\right)
\end{align}
Finally, the radial functions are distributed on an energy grid, 
\begin{align}
    \rho^{\nu \nu'}(R,E) &= \sum_{nk} \rho_{nk}^{\nu \nu'}(R) G \left( E- (\varepsilon_{nk} - E_F);\delta_E \right) \nonumber\\ &\times \exp \left(-\alpha_E R\right),
\end{align}
where $G(x;\delta)$ is a Gaussian of width $\delta$ centered at $x=0$. For the materials in the dataset, the $s$, $p$, and $d$ orbitals lead to six unique components of the RAD-PDOS fingerprint \add{($\nu \nu' = \{ss, sp, sd, pp, pd, dd \}$)}. The fingerprint involves some hyperparameters for which we use the values $E_{\mathrm{min}}=-10 \;\mathrm{eV}, E_{\mathrm{max}}=10 \;\mathrm{eV}, N_E = 25, \delta_E = 0.3\;\mathrm{eV}, \alpha_E = 0.2\;\mathrm{eV}^{-1}, R_{\mathrm{min}}=0, R_{\mathrm{max}}=5 \;\mathrm{Å}, N_R = 20, \delta_R = 0.25 \;\mathrm{Å}, \alpha_R = 0.33\;\mathrm{Å}^{-1}$.

In addition to the RAD-PDOS fingerprint, we consider a low-dimensional fingerprint consisting of \add{five} features, namely the PBE electronic band gap ($\varepsilon_{\mathrm{gap}}^{\mathrm{PBE}}$), crystal formation energy ($\Delta H$), density of states at the Fermi level (DOS@$E_F$), energy above the convex hull ($\Delta H_{\mathrm{hull}}$) and the total energy per atom in the unit cell. 
The \add{low}-dimensional fingerprint is used to train a "baseline" ML model that we use to benchmark the performance of the ML model based on the more involved RAD-PDOS fingerprint. Common to all the features considered is that they are obtained from a single DFT calculation and thus are much faster to compute that the phonon frequencies.
The gradient boosting model introduces several hyperparameters such as depth of the trees, learning rate, minimum loss gain to perform a split and minimum weights in tree leafs. These parameters are optimized using Bayesian optimization where a Gaussian process is fitted to the mean test ROC-AUC of a 10-fold cross-validation. The hyperparameters used here are max depth $=8$, learning rate $=0.06$, min split loss $=0$ and min weights $=0$. 
The XGBoost classification model is in fact a logistic regression model, i.e. the output of the model is a number between 0 and 1 which is interpreted as a probability. In our case, 0 (1) refers to a dynamically stable (unstable) material.
\section{Acknowledgments*}
The Center for Nanostructured Graphene (CNG) is sponsored by the Danish National Research Foundation, Project DNRF103. This project has received funding from the European Research Council (ERC) under the European Union’s Horizon 2020 research and innovation program grant agreement No 773122 (LIMA). K. S. T. is a Villum Investigator supported by VILLUM FONDEN (grant no. 37789).
\section{Data Availability Statement}
All the crystal structures and properties will be available in the C2DB \add{(https://doi.org/10.11583/DTU.1461\\6660.v1)}.
\section{Code Availability Statement*}
All DFT calculations were performed with the open source code GPAW (\url{https://wiki.fysik.dtu.dk/gpaw/}). The CBP protocol is available as part of the open source ASR (\url{https://asr.readthedocs.io/}). The code used for generating the electronic fingerprint for the ML classification model can be found here: \url{https://gitlab.com/knosgaard/electronic-structure-fingerprints}.
\section{Competing interests}
The authors declare no competing interests.
\section{Author contributions}
S.M. and K.S.T. developed the initial concept. S.M. performed the benchmark of the CBP protocol and developed the workflow and the analysis for the pushed materials. M.K.S., N.R.K and P.M.L. conducted the machine learning analysis. K.S.T. supervised the work and helped in interpretation of the results. All authors modified and discussed the paper together.
\section*{References}
\bibliographystyle{unsrt}
\bibliography{references}
%
%
%
%
\end{document}


\section{Supplementary figures}
Benchmark for the CPB protocol for 20 materials of the C2DB database. For each material the C2DB unique identifier is reported.
\begin{figure}[H]
    \centering
    \begin{minipage}{\L\textwidth}
        \centering
        \includegraphics[width=\l\textwidth]{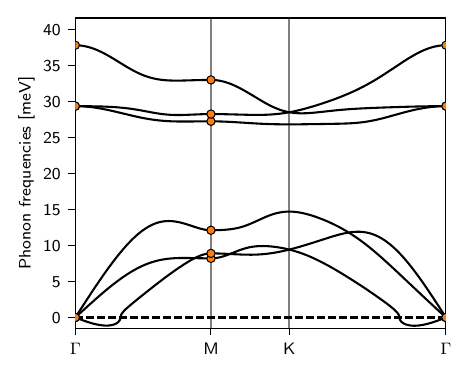}
        \caption{As2-2cfe371476fd.}
    \end{minipage}\hfill
    \begin{minipage}{\L\textwidth}
        \centering
        \includegraphics[width=\l\textwidth]{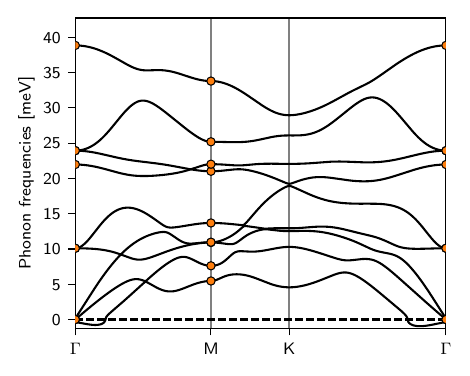}
        \caption{AsBrS-1dcd471c2288}
    \end{minipage}
\end{figure}
\vspace{-2.5mm}
\begin{figure}[H]
    \centering
    \begin{minipage}{\L\textwidth}
        \centering
        \includegraphics[width=\l\textwidth]{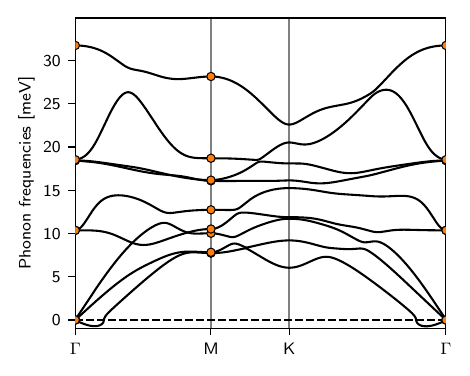}
        \caption{AsSeBr-989f469f06bd}
        \label{fig:As2}
    \end{minipage}\hfill
    \begin{minipage}{\L\textwidth}
        \centering
        \includegraphics[width=\l\textwidth]{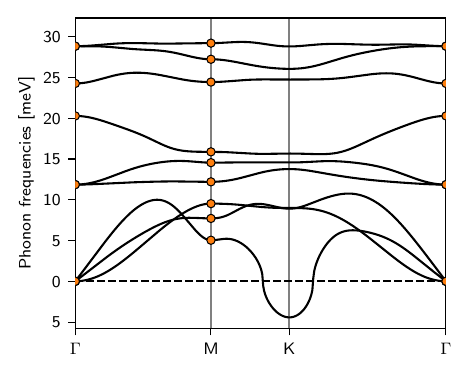}
        \caption{CoTe2-538af86b9264}
        \label{fig:AsBrS}
    \end{minipage}
\end{figure}
\vspace{-2.5mm}
\begin{figure}[H]
    \centering
    \begin{minipage}{\L\textwidth}
        \centering
        \includegraphics[width=\l\textwidth]{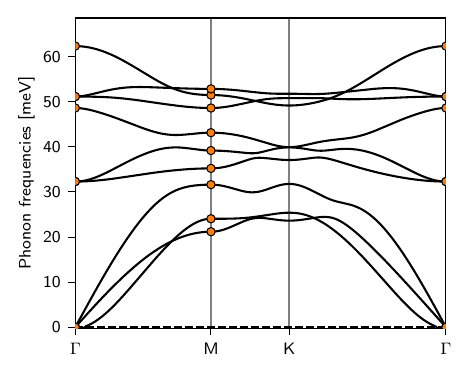}
        \caption{CrS2-c5ee5e35d2b4}
        \label{fig:As2}
    \end{minipage}\hfill
    \begin{minipage}{\L\textwidth}
        \centering
        \includegraphics[width=\l\textwidth]{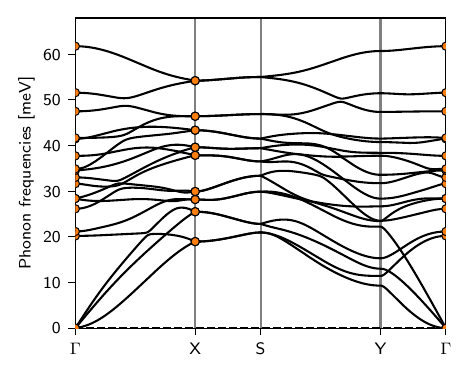}
        \caption{Cr2S4-45828f7c9596}
        \label{fig:AsBrS}
    \end{minipage}
\end{figure}
\vspace{-2.5mm}
\begin{figure}[h!]
    \centering
    \begin{minipage}{\L\textwidth}
        \centering
        \includegraphics[width=\l\textwidth]{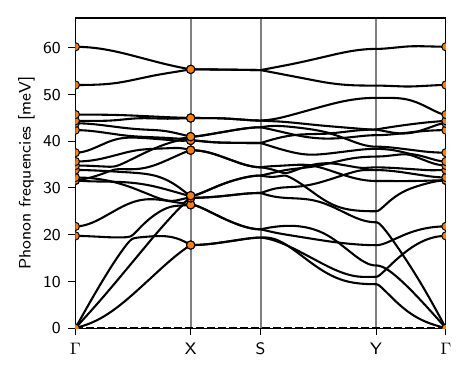}
        \caption{Fe2S4-ae77a95fa237}
        \label{fig:As2}
    \end{minipage}\hfill
    \begin{minipage}{\L\textwidth}
        \centering
        \includegraphics[width=\l\textwidth]{SI-benchmark-CBP/F2Ge2-aac9cc861c7f .pdf}
        \caption{F2Ge2-aac9cc861c7f}
        \label{fig:AsBrS}
    \end{minipage}
\end{figure}
\vspace{-2.5mm}
\begin{figure}[h!]
    \centering
    \begin{minipage}{\L\textwidth}
        \centering
        \includegraphics[width=\l\textwidth]{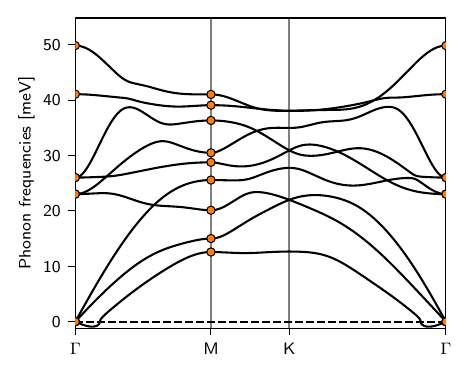}
        \caption{GeS2-6b1efcdcfd40}
        \label{fig:As2}
    \end{minipage}\hfill
    \begin{minipage}{\L\textwidth}
        \centering
        \includegraphics[width=\l\textwidth]{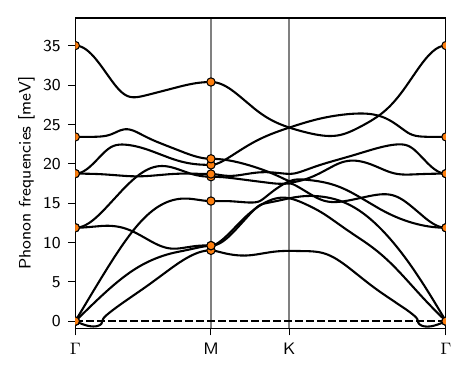}
        \caption{GeSe2-a6ec24f37a20}
        \label{fig:AsBrS}
    \end{minipage}
\end{figure}
\vspace{-2.5mm}
\begin{figure}[h!]
    \centering
    \begin{minipage}{\L\textwidth}
        \centering
        \includegraphics[width=\l\textwidth]{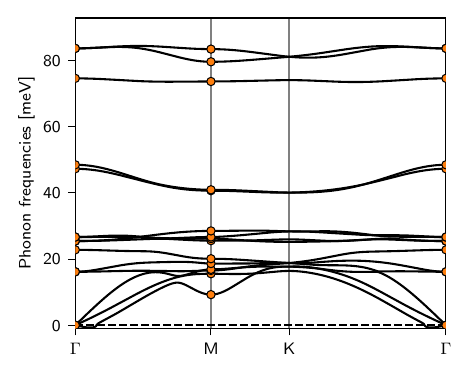}
        \caption{CF2Hf2-94a628f74b1f}
        \label{fig:As2}
    \end{minipage}\hfill
    \begin{minipage}{\L\textwidth}
        \centering
        \includegraphics[width=\l\textwidth]{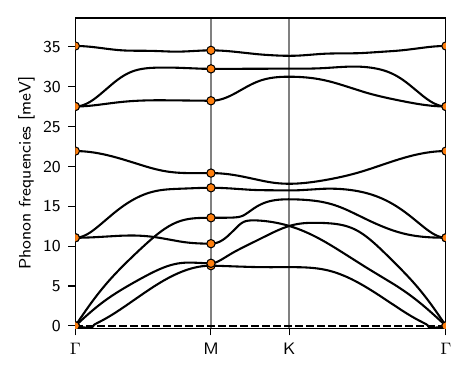}
        \caption{HfSTe-6da5c5b7dd23}
        \label{fig:AsBrS}
    \end{minipage}
\end{figure}
\begin{figure}[h!]
    \centering
    \begin{minipage}{\L\textwidth}
        \centering
        \includegraphics[width=\l\textwidth]{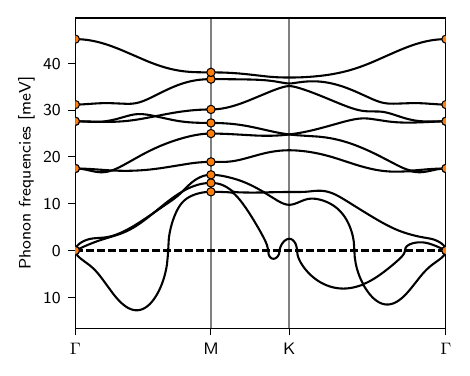}
        \caption{NbSSe-91a283a2d283}
        \label{fig:As2}
    \end{minipage}\hfill
    \begin{minipage}{\L\textwidth}
        \centering
        \includegraphics[width=\l\textwidth]{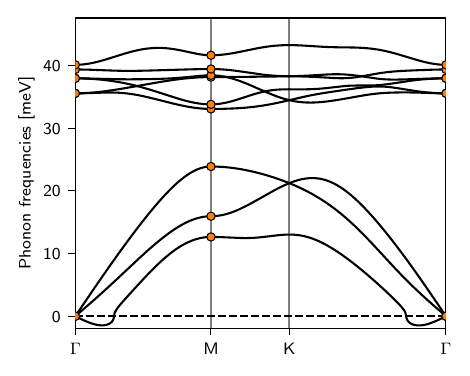}
        \caption{PtS2-77caffd1d3ed}
        \label{fig:AsBrS}
    \end{minipage}
\end{figure}
\begin{figure}[h!]
    \centering
    \begin{minipage}{\L\textwidth}
        \centering
        \includegraphics[width=\l\textwidth]{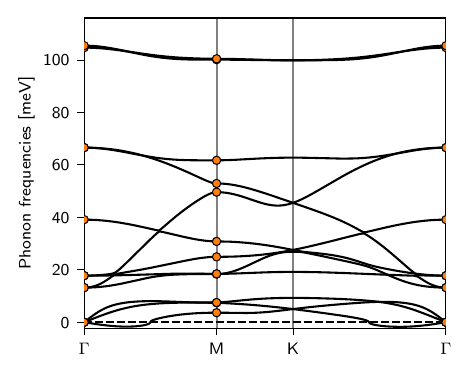}
        \caption{F2Si2-aa042bf2fe06}
        \label{fig:As2}
    \end{minipage}\hfill
    \begin{minipage}{\L\textwidth}
        \centering
        \includegraphics[width=\l\textwidth]{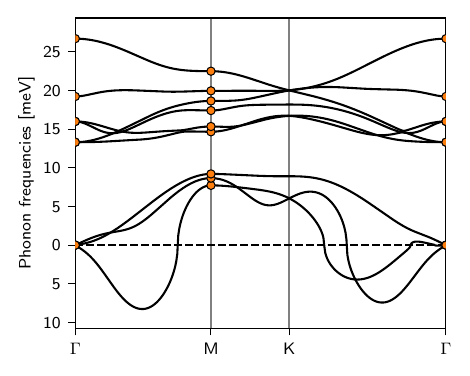}
        \caption{TaTe2-2e05715f5b75}
        \label{fig:AsBrS}
    \end{minipage}
\end{figure}
\begin{figure}[h!]
    \centering
    \begin{minipage}{\L\textwidth}
        \centering
        \includegraphics[width=\l\textwidth]{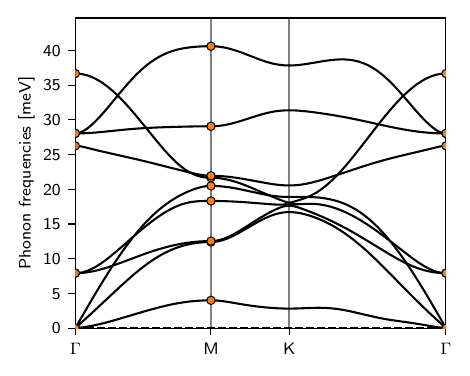}
        \caption{TiSe2-509ef368050d}
        \label{fig:As2}
    \end{minipage}\hfill
    \begin{minipage}{\L\textwidth}
        \centering
        \includegraphics[width=\l\textwidth]{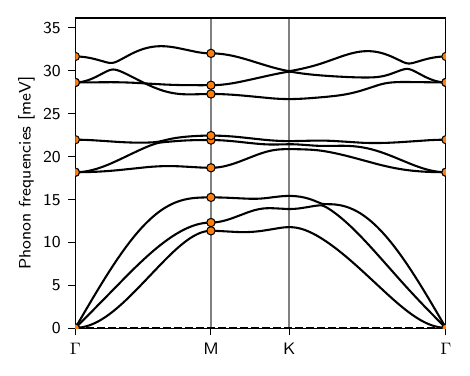}
        \caption{BrClV-878148656c6a}
        \label{fig:AsBrS}
    \end{minipage}
\end{figure}
\begin{figure}[h!]
    \centering
    \begin{minipage}{\L\textwidth}
        \centering
        \includegraphics[width=\l\textwidth]{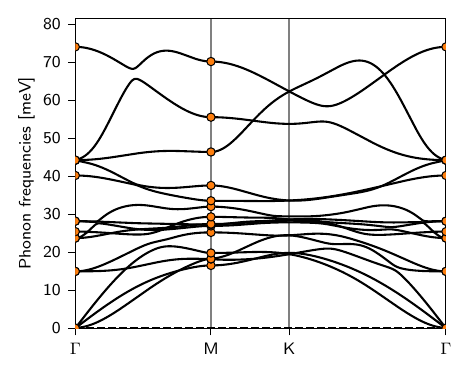}
        \caption{CF2Y2-c1e2c223ffeb}
        \label{fig:As2}
    \end{minipage}\hfill
    \begin{minipage}{\L\textwidth}
        \centering
        \includegraphics[width=\l\textwidth]{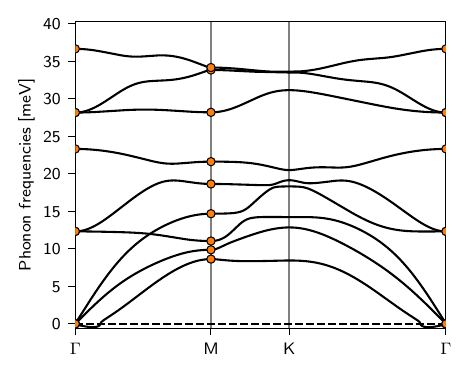}
        \caption{STeZr-3f7c4bfed2d9}
        \label{fig:AsBrS}
    \end{minipage}
\end{figure}
\clearpage
\section{Supplementary methods}
\begin{figure*}[h!]
    \centering
    \includegraphics[width=0.45\textwidth]{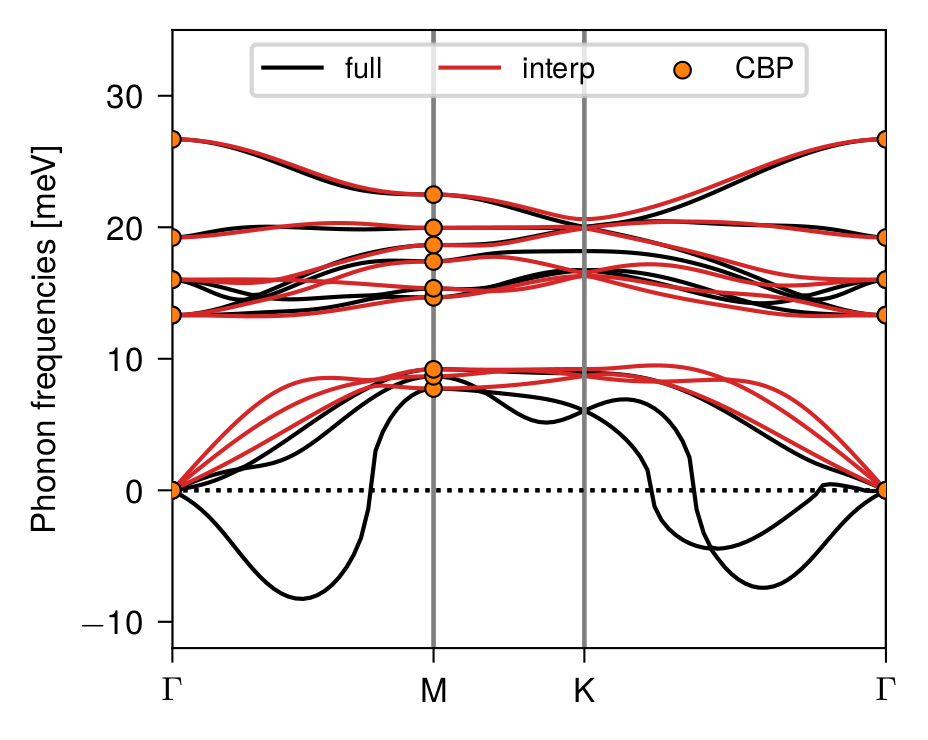}
    \caption{\textbf{Comparison between inteporlating phonon frequencies and the full phonons calculation.} Interpolating the phonons frequencies from the CBP protcol (red) is not able to catch the instability at intermediate point of the BZ (black) for the false-positive 1T-TaTe2 (uid=2e05715f5b75).}
    \label{fig:matgap}
\end{figure*}
\begin{figure*}[h!]
    \centering
    \includegraphics[width=0.85\textwidth]{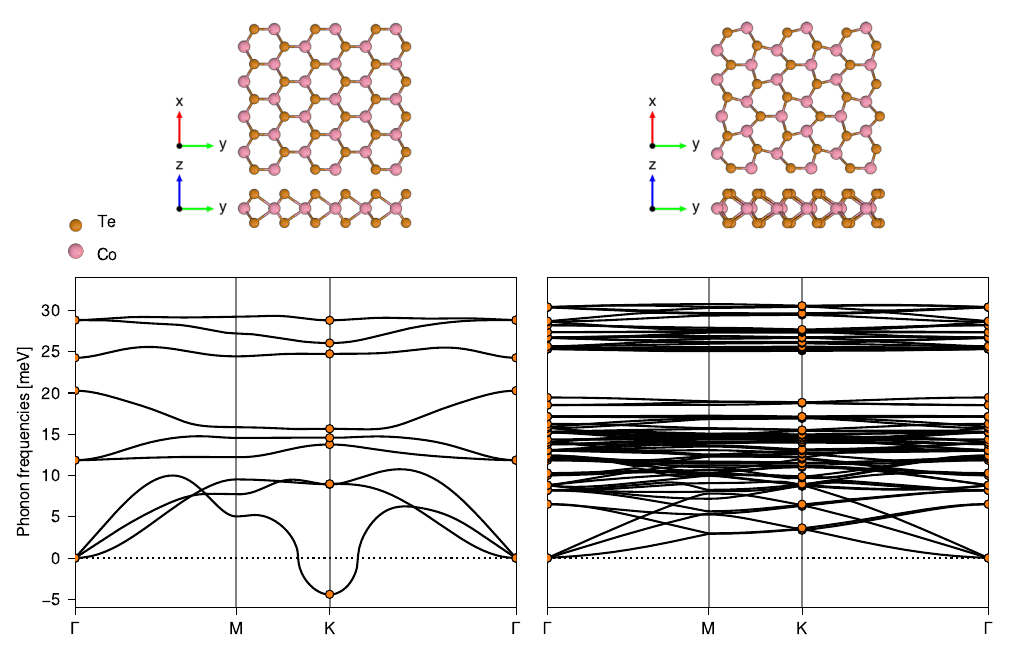}
    \caption{\textbf{CPB protocol in a 3x3 supercell.} Applying the protocol in a 3x3 supercell is possible to catch instability at the K point of the hexagonal BZ of 2H-CoTe2 (left). Displacing at K in a 3x3 supercell along the imaginary mode a stable structure is obtained (right).}
    \label{fig:matgap}
\end{figure*}
\begin{figure*}[h!]
    \centering
    \includegraphics[width=1.\textwidth]{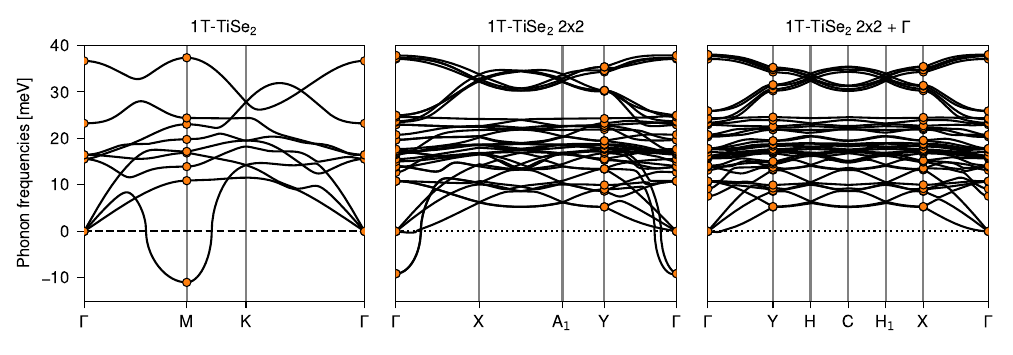}
    \caption{\textbf{The CBP protcol applied many times.} After displacing at (1/2,1/2,0) in a 2x2 supercell along the imaginary mode in the phonons bands structure of 1T-TiSe2 (left), the structure is still unstable at $\Gamma$ (center). Displacing again along the unstable mode at $\Gamma$ a stable structure is obtained (right).}
    \label{fig:matgap}
\end{figure*}